\documentstyle[prd,aps,epsf]{revtex}
\begin{document}
\draft

\input epsf \renewcommand{\topfraction}{0.8}

\title{Instantons from Low Energy String Actions.}
\author{ P. M. Saffin$^{1}$\thanks{E-mail: p.m.saffin@sussex.ac.uk}\&
          Anupam Mazumdar$^{2}$\thanks{E-mail: anupamm@star.cpes.susx.ac.uk}\&
          E. J. Copeland$^{1}$\thanks{E-mail: E.J.Copeland@sussex.ac.uk}  }
\address{$^{1}$Centre for Theoretical Physics, 
               University of Sussex,
               Falmer, 
               Brighton,
               U.K.
               BN1 9QJ\\ 
         $^{2}$Astronomy Centre, 
               University of Sussex, 
               Falmer, 
               Brighton,
               U. K.
               BN1 9QJ}
\date{\today}
\maketitle
\begin{abstract}
We look for instanton solutions in a class of two scalar field 
gravity models, which 
includes the low energy string action in four 
dimensions. In models where the matter field 
has a potential with a false vacuum, we find that 
non-singular instantons exist as long as the Dilaton field found in string theory 
has a potential with a minimum, 
and provide an example of such an instanton. The class of singular 
instanton solutions are also examined, and we find that depending on the 
parameter values, the volume factor of the Euclidean region does not 
always vanish fast enough at the singularity to make the action finite.   

\end{abstract}



\section{ Introduction }
In a fascinating letter, Hawking and Turok \cite{Hawking} 
have recently discovered a new instanton solution which they 
argue corresponds to the nucleation of a Universe undergoing 
a period of open inflation \cite{TR}, independent of the need for a 
period of false vacuum dominated inflation. Although the 
interpretation of the result has been questioned by a number of authors 
\cite{Linde,Vilenkin}, the fact that the instanton solution exists has 
generated a great deal of excitement \cite{Bousso,Unruh,Garriga,Turok}.  

The instantons, which are non trivial solutions to the euclidean equations
of motion, are interpreted as the spontaneous quantum nucleation of a 
spacetime
with some distribution of fields defined on it whose future evolution is
determined by the analytic continuation of the euclidean solution. 
A natural definition of cosmic time coincides with moving from 
one hyperbolic spatial section to another which leads to the interpretation 
of the 
solution as an open universe emerging from quantum tunneling.

Previous studies of instanton solutions possessing either  
$0(4)$, \cite{Coleman} or $0(5)$, \cite{Moss}
invariances have assumed a constant gravitational
coupling. Although today there exist tight constraints on the 
allowed spatial and temporal variation of Newtons constant \cite{Will},
these constraints could have been relaxed in the past. In other words 
we know that Einstein's theory of gravity is an
effective theory, and the correct theory could well have allowed 
for the evolution of the fundamental coupling constants. 
In this letter we treat the gravitational coupling to matter as one such 
dynamical variable. Such theories have been well studied 
\cite{JBD} and are known as scalar-tensor
theories. Among the variants is the Jordan,
Brans and Dicke (JBD) theory \cite{JBD} which treats the coupling of a scalar 
field 
to the metric as a constant. Perhaps of greater significance though 
is the fact that the gravitational
lagrangian inspired by low energy effective action for bosonic string theory 
\cite{Frad} can also be regarded as a variant, where here
the coupling constant determines the strength between the dilaton and graviton 
degrees of freedom. It is certainly worth while investigating the role 
of the Dilaton field in determining the corresponding Instanton solutions. 

In this letter, we shall show that the presence of a varying dilaton field, 
can lead to both singular and non-singular 
instantons. Moreover we shall see that depending on the parameters 
of the model, unless we  
invoke a potential in the dilaton sector the euclidean
action may become divergent. 

In \S 2 we  consider the form of the gravitational action in the Einstein
frame of reference (related to the string frame by a conformal 
transformation) and detail the type of solutions we shall be investigating,
based on an ansatz for the metric and field distributions.
We present an argument which describes how regular bubble solutions
are expected to exist when the matter field possesses a 
metastable vacuum, and show that in this case 
the dilaton is required to have a potential of its own which has some
minimum. In 
\S 3 we address the potentially singular behaviour of the instantons 
and show that
the euclidean action can be well defined (for singular instantons) 
or divergent depending on the values of the parameters in
the model being looked at. This is then discussed in the context of 
instantons arising out of string theory. 


\section{The Model}
The starting point of our analysis will be the action given by
\begin{eqnarray}
\label{BD2}
S&=&\int d^{4}x\sqrt{-g} 
       \left[-\frac{1}{2\kappa^2}R
             +\frac{1}{2}(\nabla\phi)^{2}
             +\frac{1}{2}e^{-\gamma \kappa \phi} (\nabla\sigma)^{2}
             - e^{-\beta\kappa\phi}V(\sigma) - U(\phi) \right],
\end{eqnarray}
where $\kappa^2 \equiv 8\pi G$ and $G$ is Newton's constant. 
By choosing specific values for $\beta$ and $\gamma$ we return to some
more familiar models \cite{Maeda}.
In particular, $\phi$ is related to the Brans-Dicke field in the JBD
model or the dilaton in the low energy, dimensionally reduced superstring
case. Here we shall refer to $\phi$ as the dilaton and $\sigma$ as the matter 
field. 
As is common practice in finding instantons we restrict our
attention to solutions of the euclidean equations
of motion which have an O(4) symmetry. That is to say, we impose a 
euclidean
metric of the form
\begin{eqnarray}
\label{MET}
ds^2 &=& d\xi^2+b(\xi)^2 \left(d\psi^2+sin^2 (\psi) d\Omega^2_{(2)} \right)
\end{eqnarray}
and require that the fields $\phi$ and $\sigma$ depend only on the radial 
coordinate $\xi$.
By doing this we will of course miss solutions, but this ansatz has the 
quality that instantons with this symmetry are expected to have the lowest 
euclidean
action,
although this has only been proved for the flat space case \cite{Coleman}.
The field equations and gravitational constraint for the euclidean action are 
found to be
\begin{eqnarray}
\label{MOT1}
\phi^{\prime\prime} + 3\frac{b^{\prime}}{b}\phi^{\prime} &=&
     -\frac{1}{2}\gamma e^{-\gamma\phi}\sigma^{\prime 2} 
     - \beta e^{-\beta\phi}V(\sigma) 
     + \frac{\partial U}{\partial\phi}  \, ; \\
\label{MOT2}
\sigma^{\prime\prime} + 3\frac{b^{\prime}}{b}\sigma^{\prime} &=&
      \gamma \phi^{\prime} \sigma^{\prime} 
     + e^{-(\beta -\gamma)\phi}\frac{\partial V}{\partial \sigma} \, ;\\
\label{MOT3}
b^{\prime2} &=& 1 
               + \frac{1}{3}b^{2} \left[\frac{1}{2}\phi^{\prime 2}
                      + \frac{1}{2}e^{-\gamma \phi}\sigma^{\prime 2} 
                      - e^{-\beta\phi}V(\sigma)  
                      - U(\phi) \right],
\end{eqnarray}
where $\phi^{\prime} \equiv d\phi /d\xi$ etc, and 
we have rescaled the functions and coordinates by 
$\kappa$ to remove it from the problem.
These equations may be used to generate the second order
evolution equation for $b(\xi)$
\begin{eqnarray}
\label{MOT4}
b^{\prime\prime}/b&=&-\frac{1}{3}\left[ \phi^{\prime 2}+e^{-\gamma \phi} 
\sigma^{\prime 2} 
                                     +e^{-\beta \phi} V(\sigma) + 
U(\phi)\right].
\end{eqnarray}
From this we see that $b(\xi)$ is convex as the right hand side of 
Eq.~(\ref{MOT4})
is negative semi definite. Taking $b(\xi)$ as positive in the region of
interest shows that it will vanish at two points, it then looks like a deformed
sine function with a maxima at some $\xi_{max}$, meaning that Eq.~(\ref{MET}) 
represents a 
deformed four sphere. The vanishing
of $b(\xi)$ then occurs at the poles of this deformed $S^4$ and are denoted by 
$\xi_S(=0)$ and
$\xi_N$.
If we want to find regular finite solutions to 
Eqs.~(\ref{MOT1}-\ref{MOT3}) then the 
boundary conditions
must be such that $\phi^{\prime}(\xi_N)=\phi^{\prime}(\xi_S)=0$ and
$\sigma^{\prime}(\xi_N)=\sigma^{\prime}(\xi_S)=0$, otherwise the $b^{\prime}/b$ 
term 
will generate a divergent solution. Applying this condition
to $\phi(\xi)$ we see immediately that without a dilaton potential, $U(\phi)$ 
the right hand side of Eq.~(\ref{MOT1}) is negative
semi--definite and so as we evolve from $\xi_S=0$ 
this drives \mbox{$\phi(\xi \rightarrow \xi_N) \rightarrow -\infty$}. The 
limiting behavior
at this singular point leads to a curvature singularity and, depending on 
the value of $\beta$, can
also cause the euclidean action to diverge unlike the instantons considered 
by Hawking and Turok \cite{Hawking}. 
If we are to find
instantons with non singular profiles then we take $V(\sigma)$ to have a local 
minima
at $\sigma=0$ and global one at $\sigma=\eta$. We have seen that $\phi$ must 
have a potential, 
but does this mean that if $U(\phi)$ exists we can find an instanton?
Fortunately, as we will now show, $U(\phi)$ can be fairly generic. 
Following Coleman \cite{Coleman1}, we consider 
Eqs.~(\ref{MOT1}-\ref{MOT3}) as representing two particles, 
$\Sigma$ and $\Phi$ with 
positions 
$\phi$ and $\sigma$ moving in their respective potential wells $-U(\phi)$
and $-V(\sigma)$ with some coupling
to each other
and a damping term $3\frac{b^{\prime}}{b}$, which actually becomes anti damping 
when $\xi>\xi_{max}$. 
The potentials that the particles move in are represented in figures 1 and 2.
The idea now is to find values for $\phi(\xi=0)$ and $\sigma(\xi=0)$ which 
will satisfy the boundary 
conditions at the poles.
To analyze the problem we look for a self consistent set of profiles. Start 
by assuming that
a regular profile exists for $\phi(\xi)$ and look at the matter field 
$\sigma$. The 
$\sigma$ 'particle', $\Sigma$ begins
at $\sigma(\xi=0)=\sigma_i$ with zero velocity and it rolls down the potential 
$-V(\sigma)$. At $\xi=\xi_N$
it is required to have stopped, which will take an amount of time determined by 
the shape of the potential.
If $\Sigma$ starts too close to $\sigma(\xi=0)=\eta$ then it can remain almost 
stationary up to $\xi_{max}$
when the anti damping term takes hold and pushes $\sigma$ to minus infinity, 
representing an overshoot solution. In order to make $\Sigma$ have a turn 
around point before $\xi_N$ then
$\Sigma$ must start
further from $\eta$, and if it starts just above $\sigma_*$ then it will turn 
around in a timescale
governed by the harmonic oscillations due to $-V(\sigma)$ around $\sigma_*$. So 
in fact, depending on the
nature of the potential, non trivial solutions may not exist if the harmonic 
oscillator timescale is
longer than $\xi_N-\xi_S$, in which case the only instanton solution 
then available to us is 
the analogue of the 
Hawking Moss Instanton \cite{Moss} where the fields are constant,
\begin{eqnarray}
\sigma&=&\sigma_* \, ;\\
\beta e^{-\beta \phi_*}V(\sigma_*)&=&\frac{\partial{U}}{\partial \phi}|_{\phi_*} \,;\\
b(\xi)&=&H^{-1}\sin(H \xi) \,;\\
H^2&=&\left[e^{-\beta \phi_*}V(\sigma_*)+U(\phi_*)\right]/3 .
\end{eqnarray}
In this case the metric represents that of an exact $S^4$, so that the symmetry 
of this solution is
extended from the assumed O(4) to O(5). If we are in the situation
where harmonic oscillations are faster than the $b(\xi)$ timescale then there 
will be a value, 
$\sigma_*<\sigma_i<\eta$, such that $\sigma^{\prime}(\xi_N)=0$. Turning our 
attention to the particle 
with position $\phi$, $\Phi$, we 
assume that $\sigma(\xi)$ has a regular form and then roll $\Phi$ in its 
background. 
The forcing terms for $\Phi$
on the right hand side of Eq.~(\ref{MOT1}) consist of the first 
two terms driving 
$\Phi$ in the negative direction and
a potential term which depends on the position of $\Phi$. If we consider the 
potential shown in figure 2 then by starting
$\phi(\xi=0)=0$, all the terms force $\phi$ negative causing 
\mbox{$\phi(\xi=\xi_N) \rightarrow -\infty$}. This is the
undershoot solution. By starting $\Phi$ sufficiently large and positive, the 
potential term can
always dominate the other forces so that \mbox{$\phi(\xi=\xi_N) \rightarrow 
+\infty$}, 
the overshoot solution. It should then be 
clear that in between the overshoot and undershoot regimes 
there is a case where 
$\phi(\xi)$ remains
finite. To obtain a qualitative picture of the form for $\phi(\xi)$ 
we note that if 
the negative forcing terms
dominate at the start, they always dominate, so that initially we must have 
$\phi(\xi)$ increasing from $\phi_i$
through to some maxima where the negative driving terms finally become
 important 
and reduce $\phi(\xi)$ to 
end up at
\mbox{$\phi(\xi=\xi_N) = \phi_f$}. We have numerically evolved the set of 
equations (\ref{MOT1}-\ref{MOT4})
for the specific choice of $\beta=2\sqrt{2}$, $\gamma=\sqrt{2}$ (which 
corresponds to $\phi$ being the
dilaton of the dimensionally reduced superstring action \cite{Maeda}) and a 
quadratic potential for
$\phi$. The resulting profiles are shown in figure 3 and confirm the heuristic 
arguments given above. This is encouraging as it appears possible to 
obtain finite instanton solutions for a range of initial conditions 
arising from the low energy string action. 


\section{Finite and singular Euclidean actions?}

One of the key results from Hawking and Turok was that 
the singular behavior of the tunneling matter field does  
not necessarily imply a divergent Euclidean action \cite{Hawking}. With 
this in mind we should not restrict our attention to finite Instanton 
configurations, and so we now 
consider the analogue singular solutions in the presence of the Dilaton.
As we are no longer requiring regular field profiles then the potential for 
$\sigma$ no longer needs
the rather specific form of a tunneling potential. In 
\cite{Hawking} the authors investigated a model which corresponds to 
Eq.~(\ref{BD2}) where $\phi=0$. They concluded 
that the resulting 
profiles for $\sigma$ and $b$ would have 
finite action, even
though $\sigma$ itself diverges. This is because 
the volume factor $b^3(\xi)$ suppresses 
the divergences in the 
integrand. By keeping the $\phi$ field we have obtained 
numerical solutions to Eqs.~(\ref{MOT2}-\ref{MOT4}) which can 
lead to either finite or 
divergent actions
depending on the parameters $\beta$ and $\gamma$, a result which on 
the face of it seems to go against the 
spirit of the rather general arguments presented in
\cite{Hawking}. However, our result can be traced back to the prescence of 
exponentials in the equations
of motion. Indeed we can see how the inclusion 
of such terms in the model of \cite{Hawking} may also lead to a 
divergent action. To 
see this we drop $\phi$ from the model and consider the nature of the 
singularity at $\xi_N$. 
It is expected to be dominated
by the derivatives of the profiles (backed up by numerical results), so 
that we quickly arrive at,
\begin{eqnarray}
\label{HTsing}
\sigma(\xi\rightarrow\xi_N)&\rightarrow& \ln{(\xi-\xi_N)^{\sqrt{2/3}}}\,;\\
b(\xi\rightarrow\xi_N)&\rightarrow& (\xi-\xi_N)^{1/3}.
\end{eqnarray}
The euclidean action is given by $S_e=-2\pi^2\int d\xi b^3(\xi) V(\sigma)$, 
such that its integrand for an exponential potential 
\mbox{$V(\sigma)=\exp(-A \sigma)$}, rather than power law, becomes 
$\xi^{1-A\sqrt{2/3}}$. For this to be integrable we require 
$1-A\sqrt{2/3} > -1$, corresponding to $A<\sqrt{6}$. As 
mentioned earlier, a detailed numerical 
analysis of the system of evolution
equations has borne out this inequality. 

What is the potential significance of this result? 
In low energy effective actions of string theory, the dilaton 
naturally has an exponential type potential when 
non perturbative features such as gaugino condensates are included 
\cite{gaugino}. It could be that the only allowed instanton 
solutions (i.e. finite action) are those that correspond to   
non singular profiles, which in turn require the  
introduction of a tunneling field. An analysis of the singularity in the model 
with both a dilaton $\phi$ and matter field 
$\sigma$ leads to a similar conclusion only this time the inequalities for the 
constants $\beta$ and $\gamma$ depend
on the nature of the potentials $V(\sigma)$ , $U(\phi)$. In particular, the 
values corresponding to the 
low energy superstring effective action of $\beta=2\sqrt{2}$ and 
$\gamma=\beta/2$ require a tunneling potential
for $\sigma$ as the divergence of the profiles is too strong to be 
suppressed by
the behaviour of the scale factor.

As far as the future evolution of the fields are concerned we do not consider
the details here, they may be found
elsewhere \cite{Hawking}\cite{Unruh}\cite{ancon}. The main principle to note is that the metric
Eq.~(\ref{MET}) 
can be analytically continued
to a lorentzian metric, along a slice of the instanton where the metric 
is stationary, 
which then becomes the nucleated spacetime.
The form of the spacetime clearly depends on how we choose to slice the 
instanton but 
we shall not consider this here, rather we refer the reader to the literature 
\cite{Hawking,Bousso,Unruh,ancon}.
The analytically continued metric turns out to have singularities where 
the scale
factor vanishes however, the only curvature singularity occurs for those 
cases where
the north pole of the divergent instanton is included in the instanton slice.
The other singularities of the metric turn out to be only coordinate 
singularities and the spacetime may be extended through them. 
One important aspect of this work is related to stabilizing the dilaton. 
If we are to
take the value for $\beta$ and $\gamma$ as predicted by the low energy limit 
of superstring
theory, then the singular instantons appear to be too singular to actually 
nucleate as their
action diverges. This implies that we are only allowed the regular 
instantons which, by the 
above arguments, will lead to a dilaton profile nucleated near its minimum 
and allow for the dilaton to be stabilized with a finite vacuum 
expectation value.

\section{Conclusions}
We have looked at a model of gravity which contains both the low
energy, dimensionally reduced
superstring action and the JBD theory. We have found that in the 
Einstein frame, regular instanton 
profiles exist for a class 
of dilaton potentials which possess a minimum 
when the matter field has a tunneling potential.  
Moreover, if the dilaton has 
no potential this will lead to instantons with divergent profiles. The 
finiteness of the euclidean action of such
solutions then depends on the parameters and potentials appearing in the 
specific model. The main interest of
these results lies in the superstring effective action, where a dilaton field 
arises naturally and the parameters
$\alpha=\beta/2=\sqrt{2}$ cause the euclidean action to diverge, an effect 
observed both numerically and using 
an analysis similar to the one given in the text. The outcome is a 
potentially new method of stabilising the dilaton.


\section*{Acknowledgement}

P.M.S. and E.J.C. are supported by PPARC fellowships, A.M. by Inlaks foundation and an
ORS award. We thank Kei-ichi Maeda, Jim Lidsey, Andrew Liddle for discussions,
and A.M. acknowledges use of the Starlink computer system at the University
of Sussex.


\begin{figure}[!h]
\centering
\mbox{\epsfxsize=350pt\epsfysize=350pt\epsffile{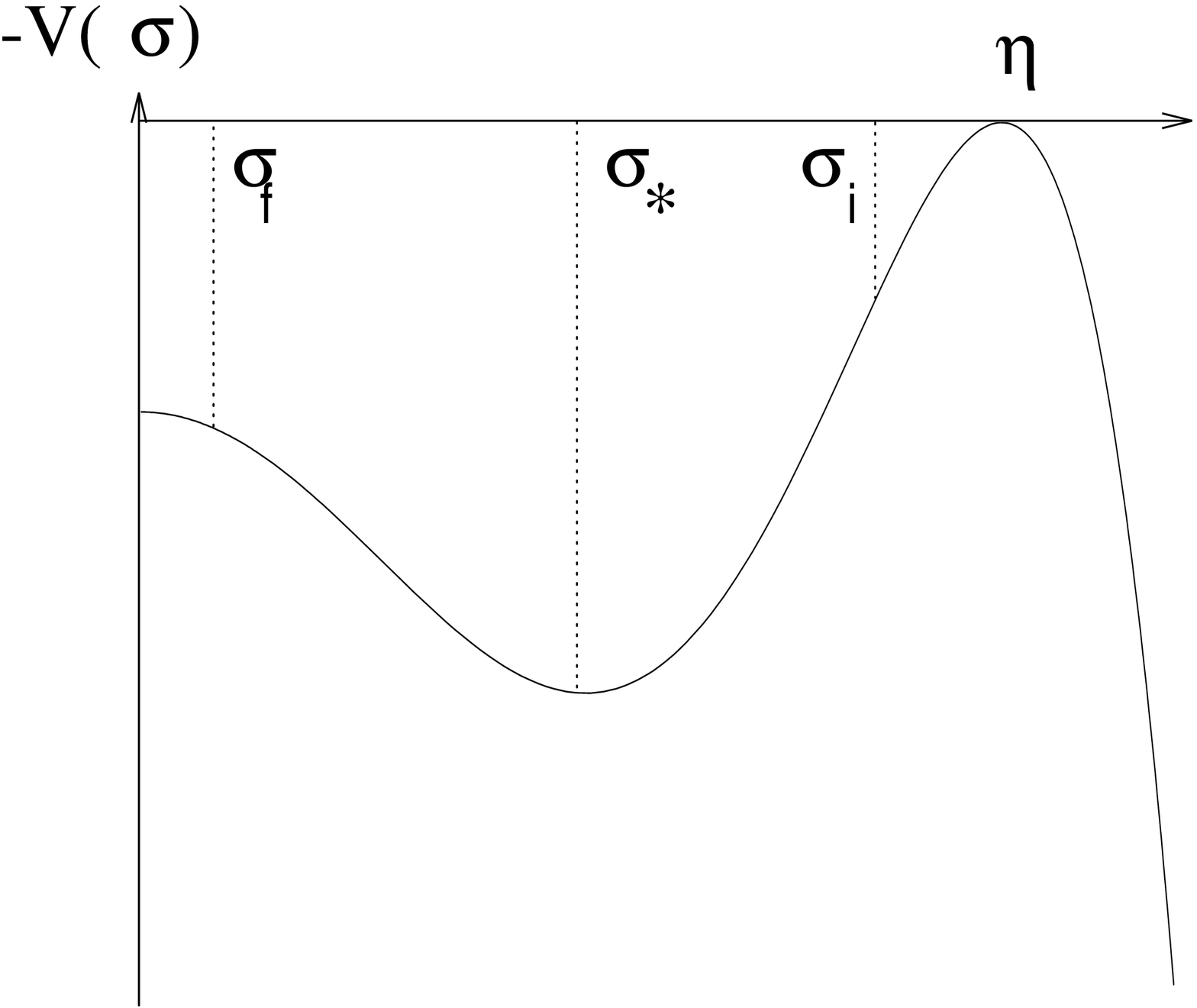}}
	\caption{
	-V($\sigma$).
	}
\end{figure}

\begin{figure}[!h]
\centering
\mbox{\epsfxsize=350pt\epsfysize=350pt\epsffile{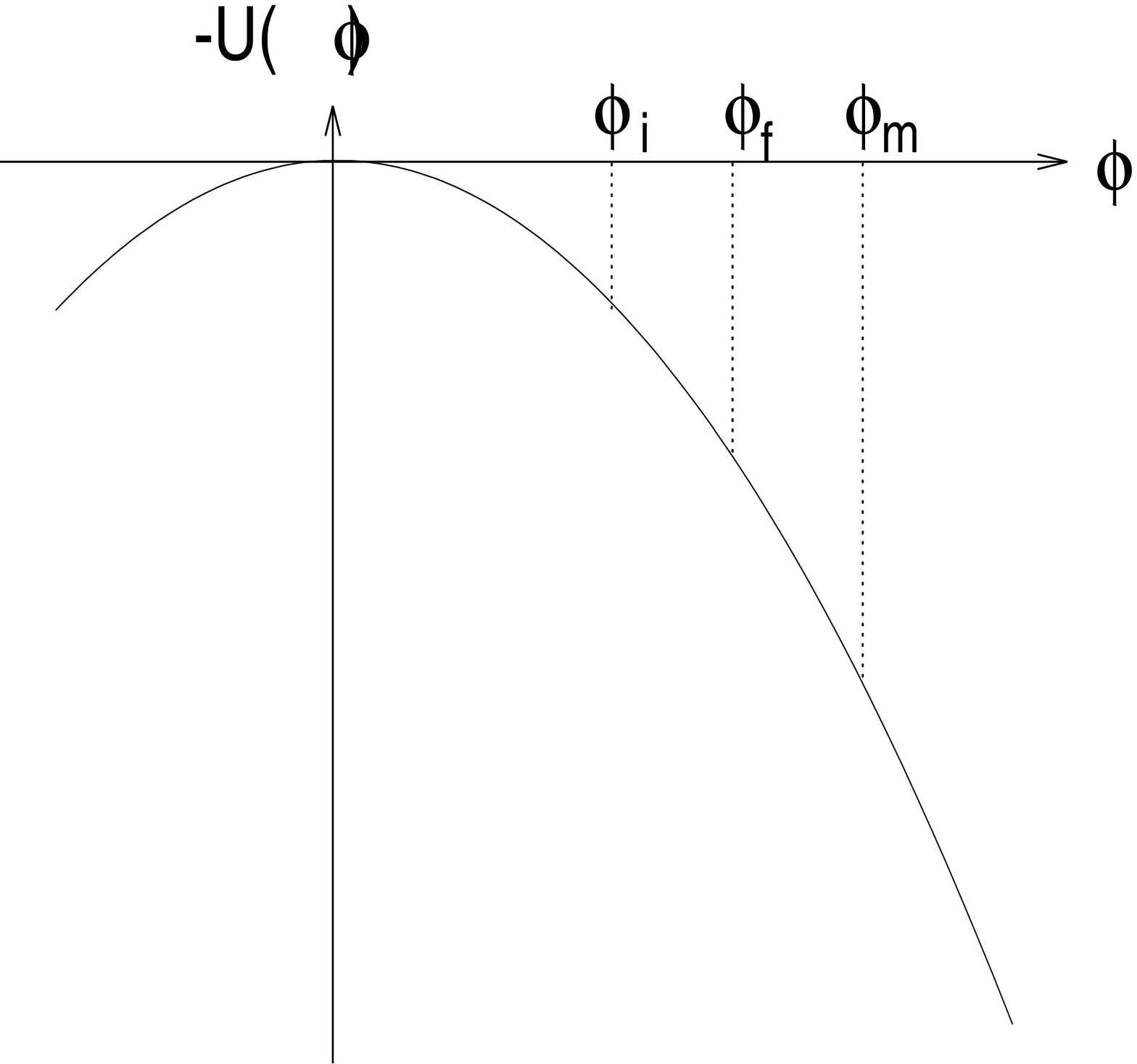}}
	\caption{
	-U($\phi$).
	}
\end{figure}

\begin{figure}[!h]
\centering
\mbox{\epsfxsize=350pt\epsfysize=350pt\epsffile{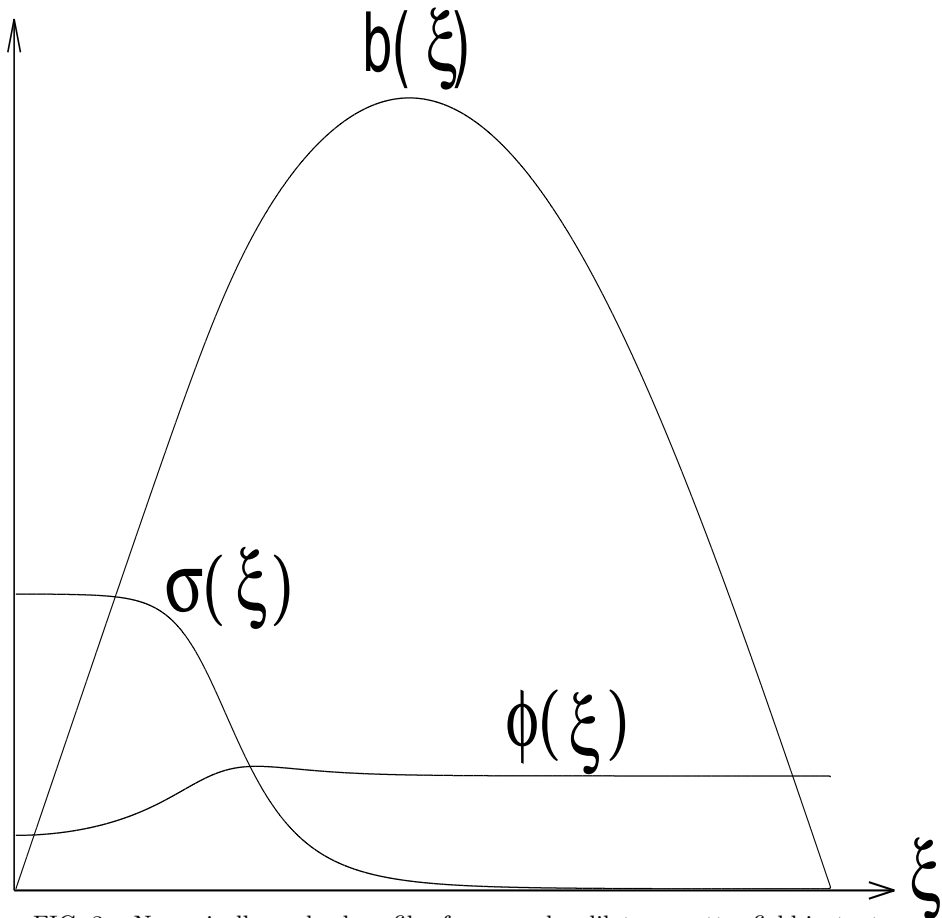}}
	\caption{
	Numerically evolved profiles for a regular dilaton, matter field
	instanton.
	}
\end{figure}


\listoffigures

\noindent Figure 1. -V($\sigma$).
\noindent Figure 2. -U($\phi$).
\noindent Figure 3. Numerically evolved profiles for a regular dilaton, matter field instanton.

\end{document}